# PERFORMANCE EVALUATION AND ENHANCEMENT OF VLAN VIA WIRELESS NETWORKS USING OPNET MODELER

Tareq Al-Khraishi and Muhannad Quwaider

Department of Computer Engineering, Jordan University of Science and Technology, Irbid , Jordan

## ABSTRACT

*A VLAN is a logical connection that allows hosts to be grouped together in the same broadcast domain, so that packets are delivered only to ports that are combined to the same VLAN. We can improve wireless network performance and save bandwidth through the characteristic VLAN network. In addition, the implementation of VLAN greatly improves wireless network security by reducing the number of hosts receiving copies of frames broadcast by switches, thus keeping hosts holding critical data on a separate VLAN. In this paper we compare wireless network with VLAN via wireless network. The proposed network is evaluated within terms of delay and average throughput using web browsing applications and file transfer in heavy traffic. The simulation was carried out using OPNET 14.5 modeler and the results show that the use of VLAN via wireless network improved performance by reducing traffic resulting in a minimized delay time. Furthermore, VLAN implementation reduces network throughput because the traffic received and transmitted has a positive relationship with throughput. Eventually, we investigated the use of adhoc routing protocols such as AODV, DSR, OLSR, TORA and GPR to improve the performance of wireless VLAN networks.*

## 1. INTRODUCTION

WLAN (Wireless Local Area Networks) allows devices to move across the network from one location to another and connect to the LAN wirelessly via radio transmission to share data and applications and other resources, without being tied to connections [1][2]. Wi-Fi (Wireless Fidelity) is referred to as the standard for WLAN communication. Today, one of the most common wireless technologies used for data transfer is the IEEE 802.11 standard, due to the need for high speed data rates many standards developed to meet customer's needs [3][4]. The table 1. Show most commonly used protocol in today's environment [5].

Table 1. Summary of Various WLAN Standards.

| Standards | RF Band | Max. Data Rate | Range |
|---|---|---|---|
| IEEE 802.11 | 2.4GHz | 2Mbs | 50 – 100m |
| IEEE 802.11b | 2.4GHz | 11Mbps | 50 – 100m |
| IEEE 802.11a | 5GHz | 54Mbps | 50 – 100m |
| IEEE 802.11g | 2.4GHz | 54Mbps | 50 – 100m |

WLAN infrastructure, as shown in Figure 1, consists of wireless stations and access points (AP) and supports many wireless stations, depending on the specification of the AP. The wireless network can be connected to the Ethernet network using a UTP cable up to 100 meters from the AP to the switch or hub [6]. Mobile stations can move while communicating and automatically search and connect to the access point device using SSID (Service Set Identifier) which is the unique name of the wireless network that matches the AP that is defined in the program, SSID keeps packets within the correct WLAN [1].

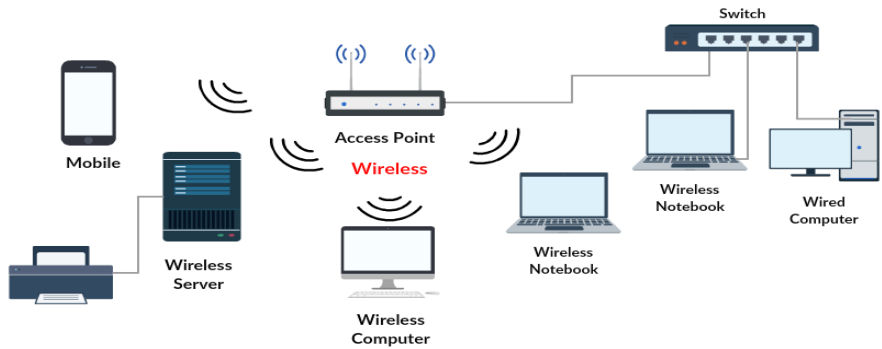

Figure 1. Wireless infrastructure diagram

There are many key performance metrics to consider when evaluating a wireless 802.11 WLAN solution, in this paper we considered throughput and delay, throughput is the rate at which the wireless LAN destination receives data successfully, while delay is the time it takes to transmit data from the source to the destination node successfully [6][7][8].

A Local Area Network (LAN) is usually defined as a broadcast domain, meaning that all connected devices are in the same physical LAN without a router. [7]. Virtual LANs that have been described regularly as a group of devices on different physical LAN segments that can communicate as if they had a common LAN segment. Switches using VLANs to divide the network into separate broadcast domains without any latency problems [8]. VLAN trunk used when connecting switches that support multiple VLANs crossing the same Ethernet connections as shown in Figure 2, the switches tag each frame sent between switches to specify the VLAN the frame belongs to. VLANs are based on the standard IEEE 802.1Q, which specifies the tagging frame format [7][9].

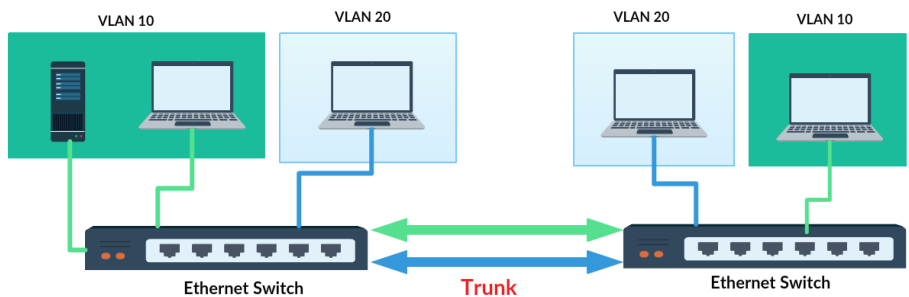

Figure 2. Network with Two VLANs

VLANs provide a number of benefits, including ease of management, decreased broadcasting and implementation of security policies [8][9]. The main advantages of VLAN are listed below:

- Performance: Network traffic is full of broadcast, so VLAN segments a large broadcast domain into a small broadcast domain that reduces unnecessary traffic in the network by sending packets only between ports that are combined to the same VLAN that reduce overhead and delay in addition to saving bandwidth.
- Organization: VLANs can be very useful to logically group hosts according to their departments or jobs that are easy to handle compared to a larger broadcast domain.
- Security: Sensitive data can be accessed by outsiders on the same network, but by creating VLAN, network security can be greatly enhanced by reducing the number of hosts receiving copies of frames that are broadcast by switches and keeping hosts holding sensitive data on a separate VLAN.
- Cost reduction: It is possible to use VLANs to create broadcast domains that cost less than costly routers.

The 802.11 communications standard defines two operating modes, Infrastructure mode and adhoc mode [10]. In infrastructure mode as shown in Figure 3, Wireless hosts can communicate with each other through access points that are typically the default mode [6]. In adhoc mode as shown in Figure 4, Without AP, wireless hosts can communicate directly with each other as a peer-to-peer network type, each host acts simultaneously as both a client and a point of access. An adhoc network is a temporary network connection created for a specific purpose in which hosts can send data directly to other hosts instead of passing through an access point [13].

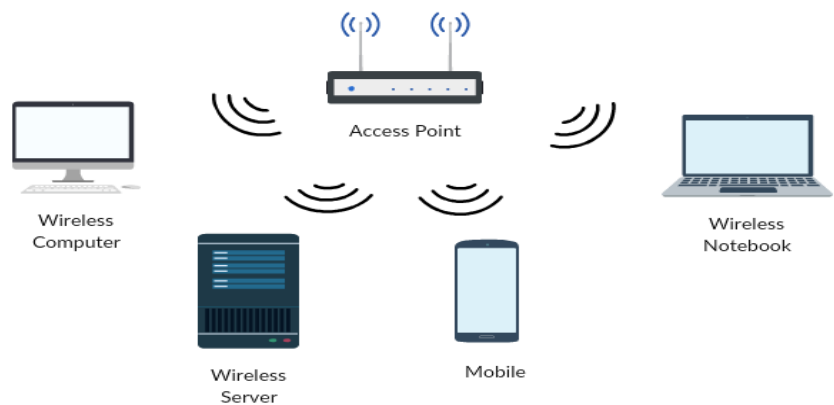

Figure 3. The 802.11 network Infrastructure operating modes

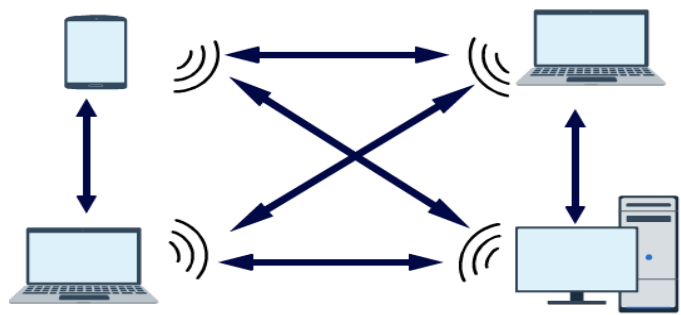

Figure 4. The 802.11 network adhoc operating modes

This paper provides enormous potential for enhancing wireless network efficiency. A brief introduction to the wireless and VLAN networks is given in this section, whereas the rest of the paper is organized as follows: Section 2 illuminates adhoc routing protocols with a focus on flat routing protocol classifications (reactive, proactive, and hybrid); Section 3 of the paper offers a brief review of related work in wireless networks and VLAN networks as well as most important previous studies on adhoc routing performance; subsequently, simulation model describes the simulation tool, simulation performance metrics, simulation setup and scenarios demonstrating three type of networks :wireless network, wireless network with VLAN and wireless VLAN network with adhoc routing protocols are given in section 4. Section 5 shows the simulation results of a comparison of all simulations performed to show the performance metrics evaluated. Finally, the conclusion is given in Section 6 based on all the work done and evaluated performance.

## 2. ADHOC ROUTING PROTOCOLS OVERVIEW

The routing protocols that have been developed for adhoc networks play an important role in affecting data transmission and network performance. Since all network nodes act as routers and participate in the discovery and maintenance of routes to other nodes on the network, each routing protocol has its own routing strategy to choose the best path between nodes. There are three main types of routing protocols for adhoc networks: proactive routing protocols, reactive routing protocols and hybrid routing protocols [13] [17].

Proactive routing is also known as table-driven routing protocols because at any time, each node in the network has one or more routes to any destination in its own routing table. [14]. Reactive routing protocols are also known as on-demand routing protocols, which are the source node to establish routes only when you have data to send, if there is no path, the protocol begins a route discovery process to find a route to the destination [17]. In hybrid routing protocols, each node acts reactively in a region that is close to it and proactively outside that region [14] [13].

A brief overview of the routing operations performed by the familiar OLSR, GRP, DSR, AODV and TORA protocols is discussed in this section. Figure 5 illustrates the routing protocols considered in this paper.

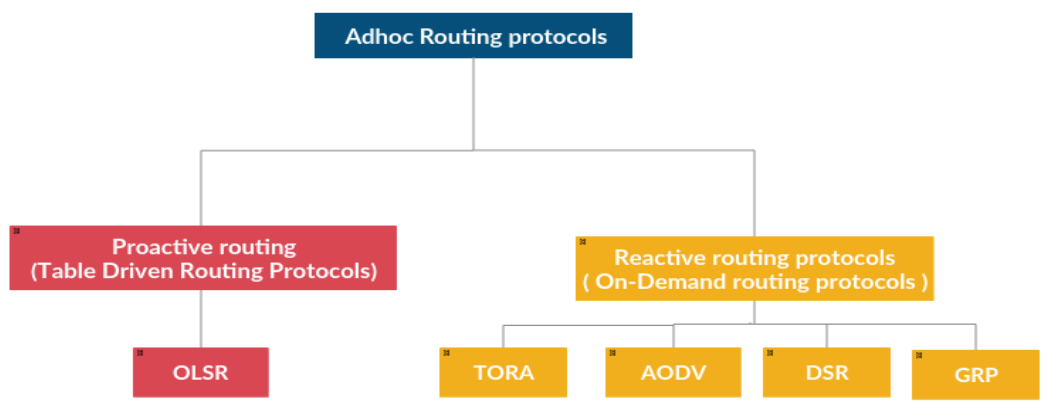

Figure 5. The adhoc routing protocols

Optimized Link State Routing (OLSR): Is a proactive protocol based on a link state algorithm [18]. The protocol's main concept is to adjust network changes without creating overhead control messages. Thus, only a group of nodes named Multi Point Relays (MPR) in the network responsible for transmitting control messages and creating link state information. Each MPR chooses to broadcast state information only between itself and the nodes that selected it [16].

Adhoc On-Demand Distance Vector Routing (AODV): Reduces the number of broadcasts by creating routes to the destination only when required, as nodes not on a selected path do not maintain routing information or participate in routing table exchanges [14]. If the source node has data to send and does not have a valid route to that destination, it initiates a path discovery process to find the destination node [15] [19].

Dynamic Source Routing (DSR): Is a simple and efficient routing protocol consisting of two mechanisms, route discovery and route maintenance [14]. The source node uses route discovery to find a route when a request arrives and inserts the paths found in the packet header. Intermediate nodes do not need to maintain up-to-date routing information other than participation in route discovery and maintenance [18][15].

Temporally Ordered Routing Algorithm (TORA): Is an on-demand routing protocol, TORA's main objective is to limit the control message in a highly dynamic mobile environment [16]. Each node must initiate a query to send the data to the destination. In essence, TORA performs three tasks: creating a route from the source to the destination, maintaining the route and erasing the route when the route is not valid.

Geographic Routing Protocol (GRP): Is classified as a proactive routing protocol. The GRP source node collects network information using the Global Positioning System (GPS) with a small amount of control overheads [17].

## 3. RELATED WORKS

VLAN can be used to create logically grouped hosts in the same domain, reduce broadcast traffic and enhance network security. It therefore offers benefits in terms of efficient use of bandwidth, performance and security. There are many authors of research on wireless and VLAN networks. The authors in [4][10][11] and [12] analyzed and evaluated the effect of the amount of traffic on the wireless network with the help of OPNET Modeler and the use of main performance metrics such as throughput, average delay and load data. The authors concluded that the delay time depends on the amount of traffic load it has.

The authors [7][8] suggests that the efficiency of LAN networks can be enhanced by using VLANs, this done by analyze and evaluate the performance of LAN and VLAN networks and measuring key performance indicators. The results showed that traffic will be decreased by increasing the number of VLANs.

Each node cooperates in a wireless adhoc network to maintain the network topology and packet transmission. We present the most important previous studies regarding the efficiency of adhoc routing.

The authors [13][17] compare the performance of two on-demand routing protocols for adhoc networks DSR and AODV. Simulation results show that DSR is more effective in a network with a small number of nodes while AODV is more effective in a higher number of nodes.

Another work [14][15] compares the performance of proactive and reactive DSDV, DSR and AODV protocols based on three performance metrics: throughput, average delay and packet delivery ratio using the NS-2 simulator. The authors conclude that DSR performs better because it has less overhead routing given the three metrics mentioned above. The authors [18] examined four adhoc routing protocols: AODV, DSR, TORA and OLSR. Throughput and delay have been used as metrics of efficiency under different scenarios. The overall results show that the OLSR performs better for a small number of nodes in the network than the reactive routing protocols DSR, AODV and TORA.

As can be seen in this section, the most relevant previous studies on wireless LAN, VLAN and adhoc routing performance are performed separately. Our work differs in that we have proposed the use of the Virtual Local Area Network (VLAN) over a wireless LAN network that can effectively constrain broadcast traffic and prevent bandwidth wastage, as well as the enforcement of security policies that are a drawback inherent in the traditional switching of wireless networks. Reduced throughput is a major drawback to using VLAN, as both short delay and high throughput are desirable in wireless networks, so we investigate this issue and improve VLAN wireless throughput performance through the use of the most widely used protocols for adhoc networks.

## 4. SIMULATION MODEL DESCRIPTION

### 4.1. Simulation Tool

Simulation design and analysis was carried out on OPNET (Optimized Network Engineering Tools) 14.5 simulator, which offers a real network scenario design, implementation and results gathering using various metrics. This paper concerns the evaluation of VLAN employment over wireless networks and examines the impact of the use of five adhoc routing protocols AODV, OLSR, DSR, TORA and GRP to improve wireless VLAN throughput performance.

### 4.2. Simulation performance metrics

For the evaluation of the network, the following two important performance metrics are considered in this paper:

- Delay: It is time for a packet to be transmitted to the destination from the source node. It is measured in seconds; better performance is expressed in lower value delay.

- Throughput: Also known as the packet delivery ratio, it is the average data packet rate that every node in the network transmits and receives successfully. It is measured per second in bits. In wireless networks, a higher value of the throughput implies better performance.

### 4.3. Simulation setup

The simulation setup Initially investigates the WLAN model with and without VLAN studies the performance of adding VLAN to the wireless network. The setup assumes that there are two servers and two switches are connecting two sections. The simulation models run with 20 nodes, each five nodes connected wirelessly to Access Point using specific BSS, randomly positioned within a square area of 1000m The MAC protocol used to be the IEEE 802.11b, which allows up to 11Mbps wireless communication [19]. Nodes (PC1-PC5) wirelessly connected to Access Point 1, (PC11-PC15) wirelessly connected to Access Point 4, (PC6-PC10) wirelessly connected to Access Point 2 and (PC16-PC20) wirelessly connected to Access Point 3, as shown in Figure 6.

Access points 1 and 3 are connected to switch 1 by 100Base-T, except that access points 2 and 4 are connected to switch 2 by 100Base-T. Ethernet Server_1 connected to Switch 1 by 100Base-T and Ethernet Server_2 connected to Switch 2 by 100Base-T.

A heavier network application traffic flow was generated with two Ethernet servers in the network being handled by each node from the respective application. The application traffic generated was as, the File Transfer Protocol (FTP) application and the Hyper Text Transfer Protocol (HTTP) applications. In addition, we also need to identify applications and profiles by adding a node for each application and then connect the workstation with the profiles.

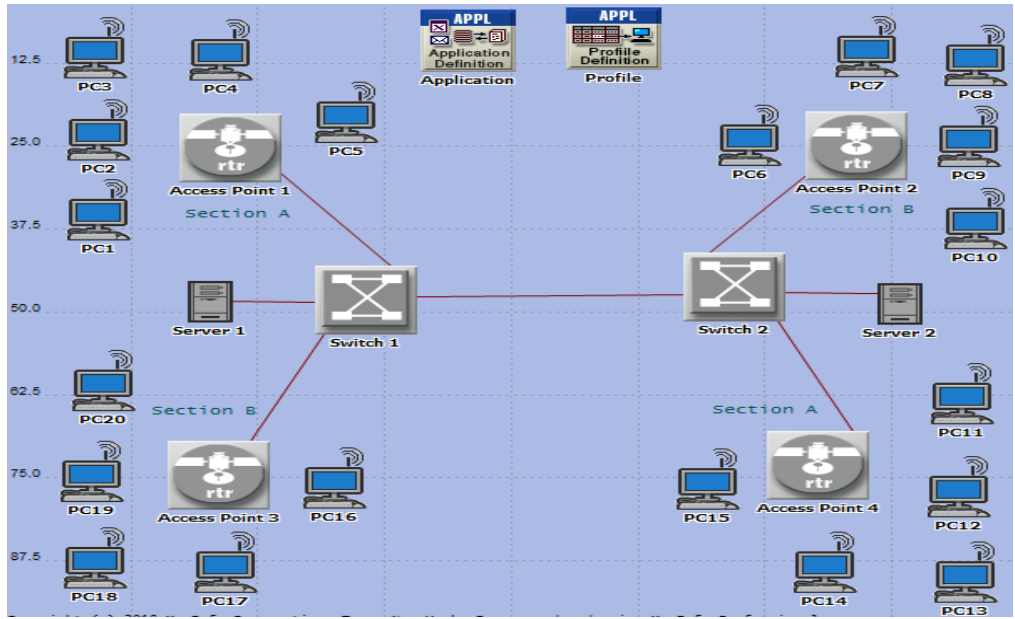

Figure 6. The Proposed wireless LAN network with 20 workstations

### 4.4. Simulation Scenarios

Regarding our purpose of simulation research, we considered a square area of 1000m here, three scenarios were created:
The first scenario (Wireless network without VLAN): This scenario considered by a wireless network as shown in Figure 6. To transmit wireless signals, we placed one access point (AP) for each section. All APs are connected by two Ethernet switches and there are two Ethernet servers that provide applications used for workstations, in this scenario there is one broadcast domain on the network, so that any workstations can communicate with two servers that have increased the wireless network throughput. Wireless LAN delay and throughput are analyzed for performance monitoring.

A second scenario (Wireless network with VLAN): In this scenario, implemented VLAN to the wireless network so the network is divided into two VLANs (VLAN10, VLAN20) as shown in Figure7. VLAN 10 consists of 10 workstations (PC1-PC5) wirelessly connected to Access Point 1 and (PC11-PC15) wirelessly connected to Access Point 4 in addition to Ethernet Server 1, while VLAN 20 consists of 10 workstations (PC6-PC10) wirelessly connected to Access Point 2 and (PC16-PC20) wirelessly connected to Access Point 3 in addition to Ethernet Server 2. VLAN trunk used when connected switches because they support multiple VLANs that cross over 100base-T Ethernet links together. Wireless LAN delay and throughput are analyzed for performance monitoring.

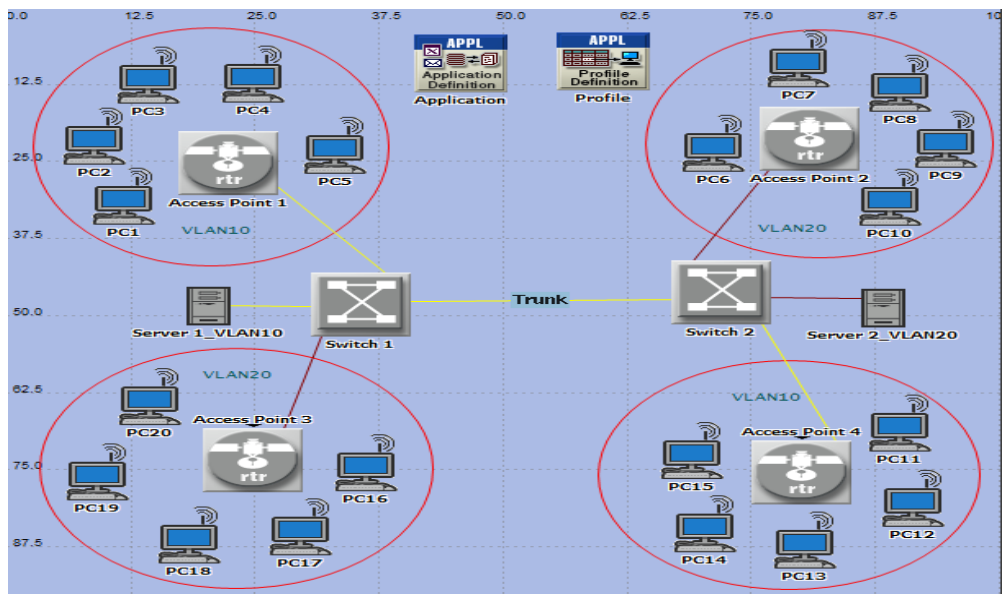

Figure 7. The Proposed wireless LAN network with two VLANs

The third scenario (Using routing protocols to improve the performance of Wireless VLAN network): This scenario focused on examining the impact of the use of five AODV, OLSR, DSR, TORA and GRP adhoc routing protocols to improve wireless VLAN throughput performance of the network. Three types of network scenarios are designed: high density, medium density and low-density networks show the effect of increasing the number of workstations on the performance of the VLAN wireless network. The low-density network consists of 10 workstations per VLAN, a medium-density network of 20 workstations per VLAN and a high-density network of 30 workstations per VLAN. Wireless LAN delay and throughput was evaluated for performance monitoring.

## 5. SIMULATION RESULTS AND DISCUSSIONS

The network was run using a wireless network and a wireless network with VLAN scenarios. The simulation parameters used to compare the two scenarios ' performance are throughput and average delay. The results of the simulation are obtained with the OPNET 14.5 modeler.

Regarding wireless delay, observe performance levels in VLAN via wireless network compared to wireless network without VLAN, as shown in Figure 8. The performance of these two networks remains very stable over the entire simulation period after the initial set-up phase. However, the traffic has been greatly reduced by using VLAN technology because the network broadcasting domain has been divided, which reduces the load on the network. In fact, VLAN improved the performance via WLAN by proving less wireless delay time.

International Journal of Wireless & Mobile Networks (IJWMN) Vol. 12, No. 3, June 2020

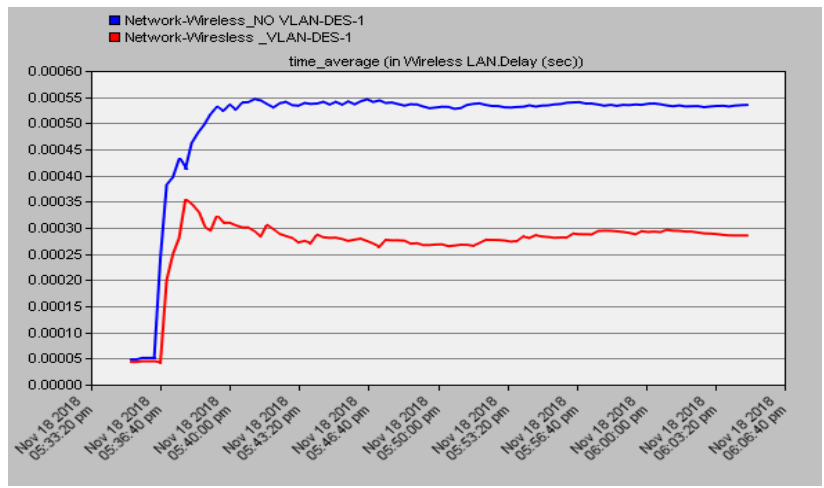

Figure 8. The wireless LAN delay in the scenario 1 and scenario 2

In terms of wireless throughput, it was found that VLAN via wireless has a lower throughput compared to the WLAN scenario as shown in Figure 9. This is because a wireless network without a VLAN scenario has higher traffic than VLAN via a wireless scenario due to a single broadcast domain. The performance of these two networks remains almost steady over the entire simulation period band after the network has converged. The received and forwarded traffic has a positive relationship with the throughput. Therefore, VLAN cannot increase the network throughput, but works on segmenting a single large network domain into multi-broadcast domains to achieve higher bandwidth utilization at a lower traffic level.

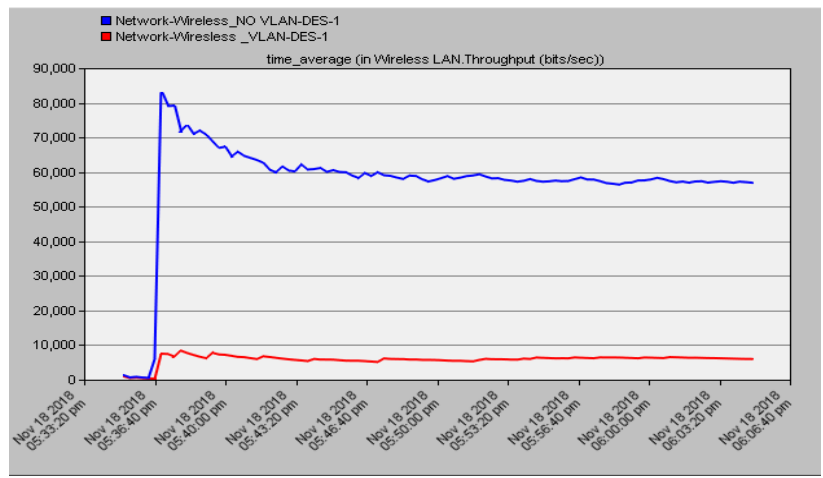

Figure 9. The wireless LAN throughput in the scenario 1 and scenario 2

Comparing the results shown in Figure 8 and Figure 9 gives a clear picture of the selection of a VLAN over a wireless network that the delay was reduced with the use of VLAN, whereas the throughput was also reduced, which is the major drawback of using VLAN, since both short delay and high throughput are desirable in wireless networks. We therefore proposed to improve the use of VLAN through the use of various adhoc routing protocols such as AODV, OLSR, DSR, TORA and GRP and to examine the impact of using these protocols to improve the performance of the wireless VLAN network throughput. Considering the results of use adhoc routing protocols as shown in Figure 10 and Figure 11. We observe that the delay can be further reduced depending on the type of routing protocol. The delay for TORA is higher because of its

23

route discovery process construction does not take place quickly and takes a lot of time to discover and decide the route of data transfer leading to potential long delays. The DSR protocol uses cached routes and often sends traffic on obsolete routes that can cause retransmissions and increase delays. OLSR has the lowest delay because OLSR is a proactive routing protocol and routes within the network are always available whenever the application layer has a traffic to transmit. We conclude that in all scenarios, OLSR showed very low delay. TORA's delay was high, AODV and GRP had an improved delay in the wireless VLAN network.

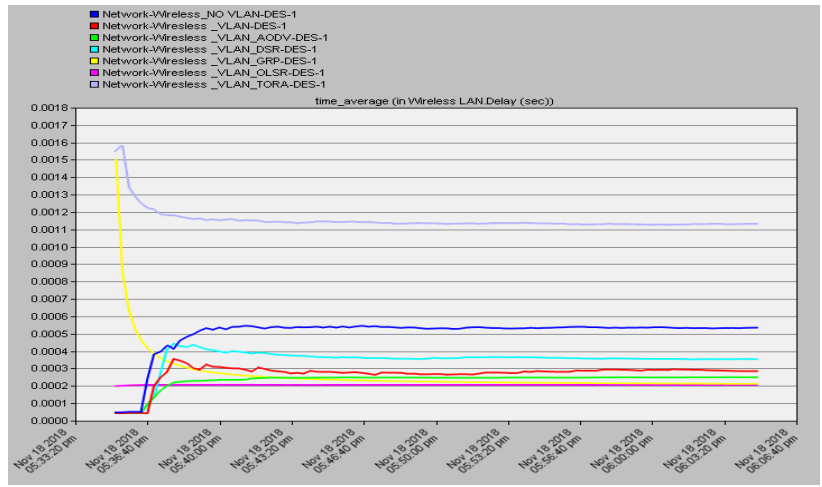

Figure 10. Comparison of Average delay in the wireless network for 20 Nodes

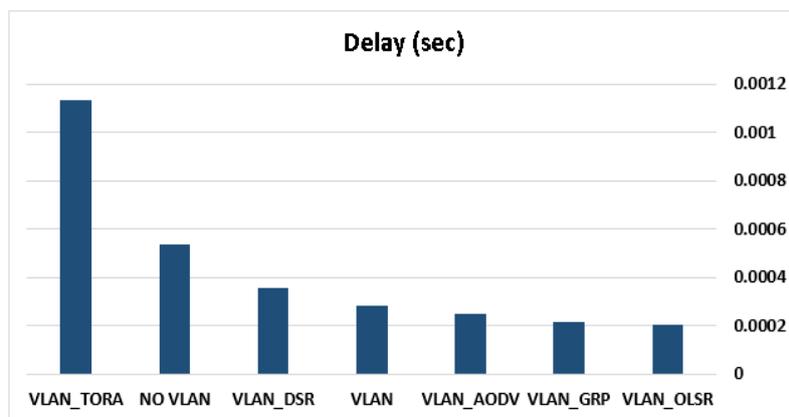

Figure 11. The wireless LAN delay for 20 Nodes

As shown in Figure 12 and Figure 13. We observe that an improvement in the throughput can be achieved through the use of the routing protocol. AODV deserves improved performance with a small number of nodes followed by DSR and OLSR.

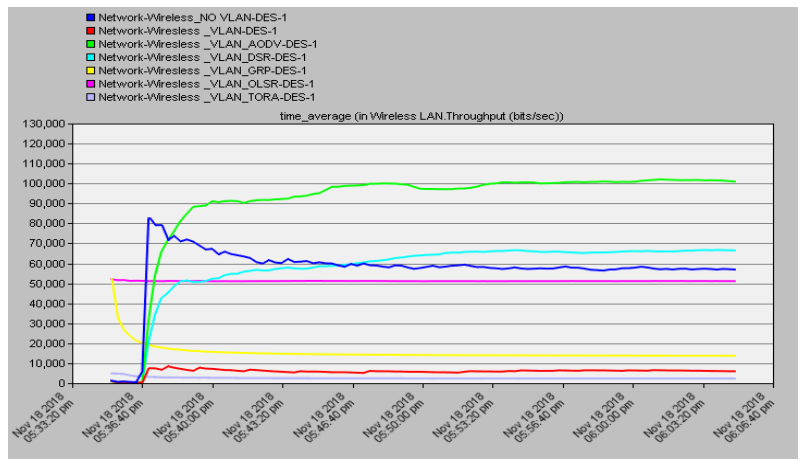

Figure 12. Comparison of throughput in the wireless network for 20 Nodes

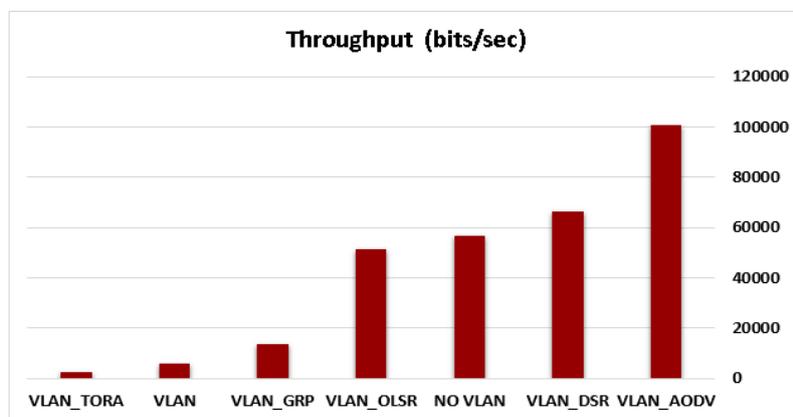

Figure 13. The wireless LAN throughput for 20 Nodes

In order to show the effect of increasing the number of workstations on the performance improvement achieved by using adhoc routing protocols on the VLAN wireless network, we return the simulation by increasing the number of nodes in each VLAN to 20 nodes and 30 nodes. In terms of throughput, with a medium density network (40 nodes) as shown in Figure 16 and Figure 17, AODV still performs better than DSR and OLSR. The rapid improvement of AODV performance is clearly seen when the number of nodes is increased. In terms of delay, as seen in Figure 14 and Figure 15, OLSR still performs well. Overall results showed that the AODV gives higher throughput as the number of nodes increases, while the OLSR gives the least delay. In addition, with regard to the AODV protocol, we note an increase in delays when the size of the network increases, so we conclude that AODV is suitable for low and medium load networks.

International Journal of Wireless & Mobile Networks (IJWMN) Vol. 12, No. 3, June 2020

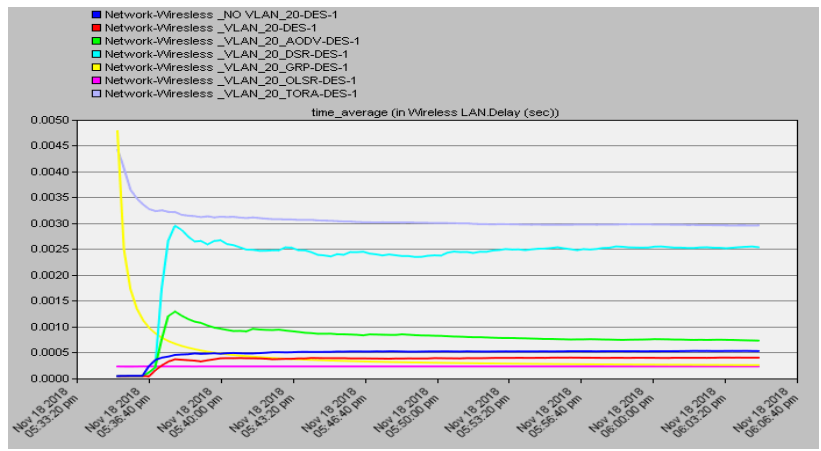

Figure 14. Comparison of Average delay in the wireless network for 40 Nodes

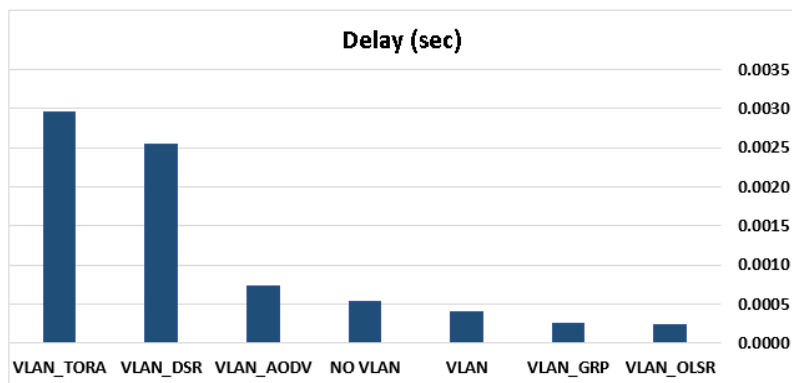

Figure 15. The wireless LAN delay for 40 Nodes

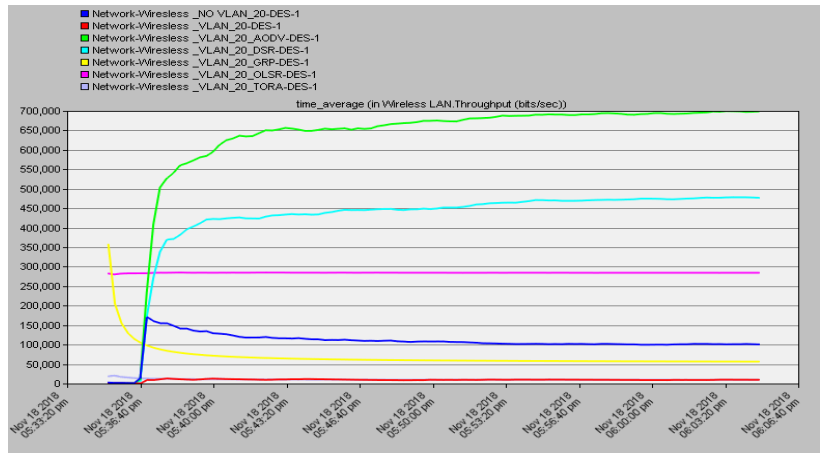

Figure 16. Comparison of throughput in the wireless network for 40 Nodes

26

International Journal of Wireless & Mobile Networks (IJWMN) Vol. 12, No. 3, June 2020

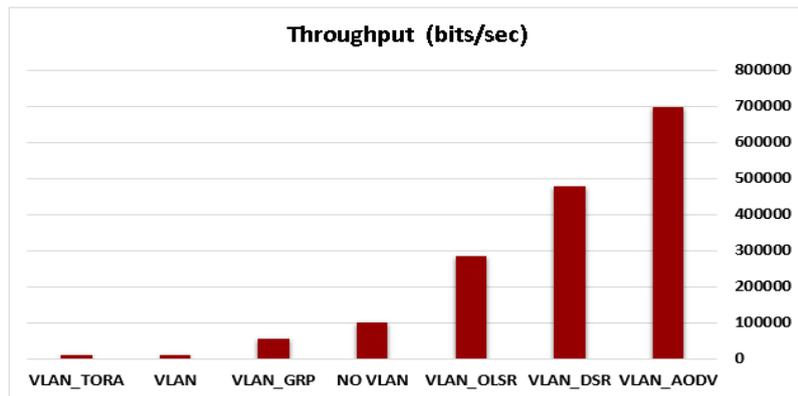

Figure 17. The wireless LAN throughput for 40 Nodes

In terms of throughput, with a high-density network (60 nodes), AODV deserves better performance, followed by DSR and OLSR, as shown in Figure 20 and Figure 21. While, as seen in Figure 18 and Figure 19, OLSR is still performing well over the delay. The decrease in AODV performance is clearly seen as the number of nodes increases.

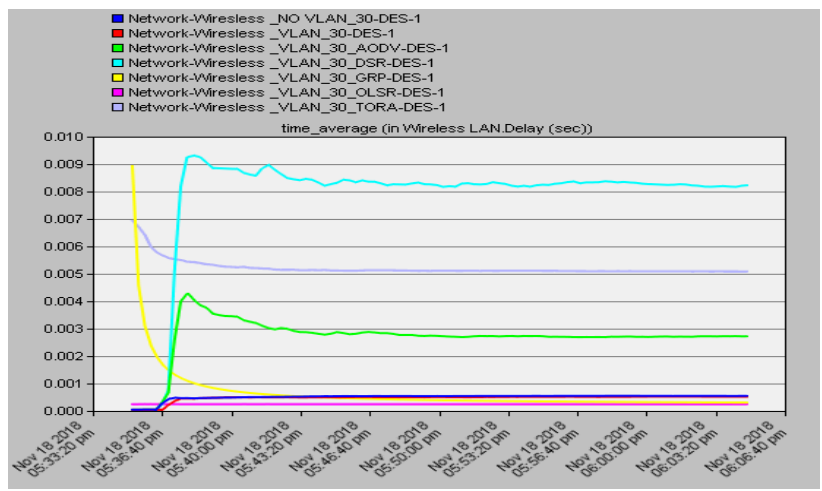

Figure 18. Comparison of Average delay in the wireless network for 60 Nodes

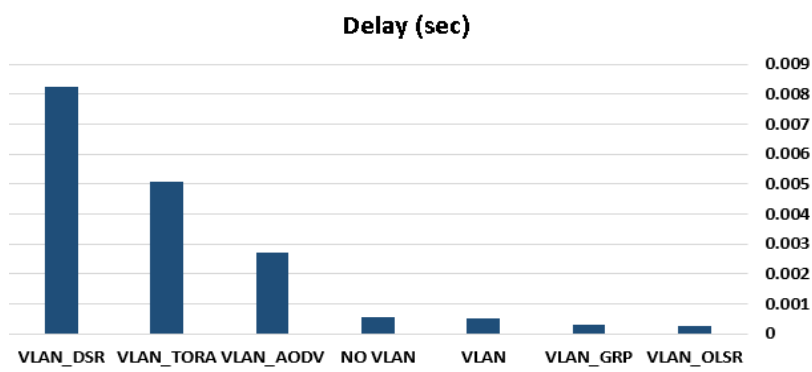

Figure 19. The wireless LAN delay for 60 Nodes

27

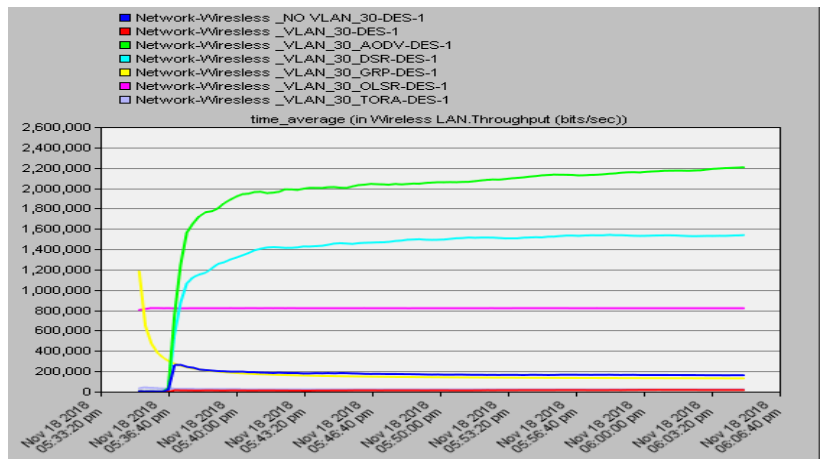

Figure 20. Comparison of throughput in the wireless network for 60 Nodes

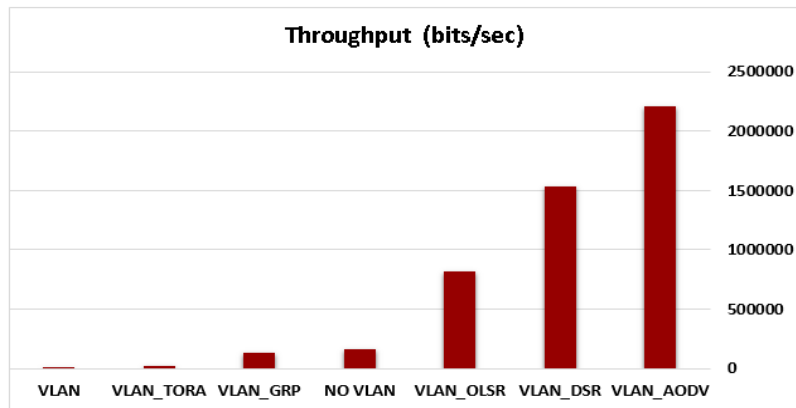

Figure 21. The wireless LAN throughput for 60 Nodes.

In conclusion, we found that OLSR shows a very low delay in all cases, while for the TORA protocol we notice that it has a big delay, also with increase of traffic the DSR has a big delay because the size of traffic has influenced the delay. As far as the AODV protocol is concerned, we notice an improvement in the throughput when the size of the network increases, but, on the other hand, an increase in the traffic size increased its delay. Finally, the results also show that the type of routing protocol is an important factor in deciding where more throughput can be improved or where more delay can be improved.

## 6. CONCLUSIONS

In this paper, we presented the use of Virtual Local Area Network (VLAN) over a wireless LAN network with the File Transfer Protocol (FTP) application and Hyper Text Transfer Protocol (HTTP) application traffic using the OPNET 14.5 simulator. Comparing performance has given a clear picture that the implementation of VLAN over a wireless network has resulted in a reduction in both delay and throughput. While it is beneficial to reduce delay, low throughput is a major drawback in wireless networks, so we used various adhoc routing protocols to improve the performance of the wireless VLAN network throughput. Overall results showed that the OLSR protocol performed better in terms of delay than other protocols, regardless of the size of the network, since OLSR is a proactive routing protocol and does not need to find routes to the destination. In a low-density network, the reactive routing protocol AODV performs better,

International Journal of Wireless & Mobile Networks (IJWMN) Vol. 12, No. 3, June 2020

whereas performance degrades when the number of nodes is increased, concluding that it is more suited for low-density network. However, while OLSR, DSR is performing well in terms of throughput, AODV is still dominant. As routing is desired for an appropriate operation of network, the network designer should select an appropriate routing protocol to suit the purpose of the network. As a result, selecting the type of routing protocol is an important factor in deciding whether throughput or delay can improve more.

## ACKNOWLEDGEMENTS

The authors would like to thank the Department of Computer Engineering, Jordan University of Science and Technology, for providing guidance and supporting this research work. We would like to express our deep appreciation and thanks to the Dean of Computer and Information Technology for helping us to receive various university facilities and services that have enabled us to successfully implement this project.

International Journal of Wireless & Mobile Networks (IJWMN) Vol. 12, No. 3, June 2020

**Authors**

**Tareq Al-Khraishi** received the B.Sc. In Computer Engineering, Al-Balqa Applied University, Jordan, 2003, and he is studying M.Sc. in Computer Engineering, Jordan University of Science and Technology (JUST), Jordan. His research interests focus on VLSI design and Nano-Technology in addition to wireless data networking, Cloud Computing and mobile ad hoc networks.

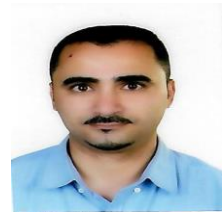

**Muhannad Quwaider** is an Associate Professor of computer engineering, college of Computer and Information Technology at Jordan University of Science and Technology (*JUST*). Dr. Quwaider earned his Ph.D. and M.S. at Michigan State University in East Lansing, USA, and his B.S. at Jordan University of Science and Technology in Irbid, Jordan. Prior to joining JUST in 2010, Dr. Quwaider was senior researcher in Networked Embedded and Wireless Systems (*NeEWS*) laboratory at the Electrical and Computer Engineering (ECE) Department of Michigan State University (MSU). Since 2012, Dr. Quwaider is Vice dean of Faculty of Computer and Information Technology, Jordan University of Science and Technology, Irbid, Jordan. His current research interests include the broad area of wireless data networking, Cloud Computing, Internet of Things, low-power network protocols, Big Data, application-specific sensor networks, wireless network security, mobile ad hoc networks, and body area network.

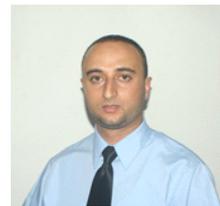